\documentclass{article}

\usepackage[utf8]{inputenc}
\usepackage{subcaption} 
\usepackage{amssymb}
\usepackage{graphicx}
\usepackage{epsfig}
\usepackage{amstext}
\usepackage{amsmath}
\usepackage{authblk}
\usepackage{color}

\usepackage{tikz}
\usetikzlibrary{arrows}
\usetikzlibrary{shapes.geometric}

\tikzset{
  treenode/.style = {align=center, inner sep=0pt, text centered,
    font=\sffamily},
  bt_l/.style = {treenode, regular polygon, regular polygon sides=3, draw=black, fill=white,
    minimum width=0.5em, minimum height=0.5em}, 
  bt_n/.style = {treenode, rectangle, draw=black, fill=black,
    minimum width=0.5em, minimum height=0.5em}
}

\title{On the Complexity of the CSG Tree Extraction Problem}
\author[1]{Markus Friedrich}
\author[2]{Pierre-Alain Fayolle}
\affil[1]{Institute for Computer Science\\
		Ludwig-Maximilians-University Munich\\
		Oettingenstr. 67\\
		80538 Munich, Germany\\
		markus.friedrich@ifi.lmu.de}
\affil[2]{Division of Information and Systems\\
		The University of Aizu\\
		Aizu-Wakamatsu City \\
		965-8580 Fukushima, Japan\\
		fayolle@u-aizu.ac.jp}
\date{}

\begin{document}

\maketitle

\begin{abstract}
In this short note, we discuss the complexity of the search space for the problem of finding a CSG expression (or CSG tree) corresponding to an input point-cloud and a list of fitted solid primitives. 
\end{abstract}

\section{Introduction}
We are interested in the problem of reconstructing a CSG expression from an unstructured point-cloud. Following \cite{fayolle2016evolutionary}, the input point-cloud is first segmented and solid geometric primitives are fitted to each segment as, for example, in \cite{friedrich2020hybrid}. Given the input point-cloud and a set of solid primitives, we need to form a CSG tree expression, involving primitives from the set of fitted primitives, and corresponding to the input point-cloud. 

Generating a CSG expression from various types of input has been the topic of multiple works, such as, for example, generating a CSG expression from a B-Rep model \cite{shapiro1991construction,shapiro1993separation,buchele2004three}, a triangle mesh \cite{du_tog2018} or a point-cloud \cite{fayolle2016evolutionary,  wu2018constructing,friedrich2018,friedrich_gecco2019}. 
Recently, the problem of generating CSG expressions from polygons or point-clouds has also attracted interest from the programming language community \cite{nandi2018,nandi2020} or from the machine learning community, with approaches relying on deep artificial neural networks \cite{sharma2018csgnet,tian2019learning,ellis2019write,sharma2020neural,walke2020learning,kania2020ucsg}. 

In this short note, we are interested in analyzing the asymptotic time complexity of the CSG tree search given a list of fitted primitives and a sampled point-cloud. 
In the following, we denote by $\Phi(P)$ a CSG tree for a primitive set $P=\{p_1, p_2, \ldots, p_{|P|}\}$. 
A CSG tree is a binary tree, its inner nodes are (Boolean) operation taken from the set of operations $O$, and its leaves are geometric primitives taken from the set of fitted primitives $P$. 
In the rest of the text, we use interchangeably the terms CSG expression and CSG tree.

\section{Enumerating CSG Trees}
To keep things simpler, we consider in this section only the binary operations: $O=\{\cup^{*},\cap^{*},-^{*}\}$ and omit the unary complement operation $\setminus^{*}$. In practice, it is always possible to simulate the complement operation ($\setminus^{*}$) from the difference operation ($-^{*}$) by adding to the list of primitives a primitive corresponding to the universe set. 

\begin{figure}[!htbp]
\centering

\begin{tikzpicture}[>=stealth',level/.style={sibling distance = 1cm/#1,
  level distance = 0.5cm}] 
\node [bt_l] {}
; 
\end{tikzpicture}

\vspace{0.5cm}

\begin{tikzpicture}[>=stealth',level/.style={sibling distance = 1cm/#1,
  level distance = 0.5cm}] 
\node [bt_n] {}
    child{ node [bt_l] {} } 
    child{ node [bt_l] {} }
; 
\end{tikzpicture}

\vspace{0.5cm}

\begin{minipage}{.15\textwidth}
\begin{tikzpicture}[>=stealth',level/.style={sibling distance = 1cm/#1,
  level distance = 0.5cm}] 
\node [bt_n] {}
    child{ node [bt_n] {} 
	child{ node [bt_l] {}}
	child{ node [bt_l] {}}
	} 
    child{ node [bt_l] {} }
; 
\end{tikzpicture}
\end{minipage}
\begin{minipage}{.15\textwidth}
\begin{tikzpicture}[>=stealth',level/.style={sibling distance = 1cm/#1,
  level distance = 0.5cm}] 
\node [bt_n] {}
    child{ node [bt_l] {} }
    child{ node [bt_n] {} 
	child{ node [bt_l] {} }
	child{ node [bt_l] {} }
	} 
; 
\end{tikzpicture}
\end{minipage}

\vspace{0.5cm}

\begin{minipage}{.15\textwidth}
\begin{tikzpicture}[>=stealth',level/.style={sibling distance = 1cm/#1,
  level distance = 0.5cm}] 
\node [bt_n] {}
    child{ node [bt_n] {} 
	child{ node [bt_n] {}
		child { node [bt_l] {}}
		child { node [bt_l] {}}
	}
	child{ node [bt_l] {}}
	} 
    child{ node [bt_l] {} }
; 
\end{tikzpicture}
\end{minipage}
\begin{minipage}{.15\textwidth}
\begin{tikzpicture}[>=stealth',level/.style={sibling distance = 1cm/#1,
  level distance = 0.5cm}] 
\node [bt_n] {}
    child{ node [bt_n] {} 
	child{ node [bt_l] {}}
	child{ node [bt_n] {}
		child { node [bt_l] {}}
		child { node [bt_l] {}}
	}
	} 
    child{ node [bt_l] {} }
; 
\end{tikzpicture}
\end{minipage}
\begin{minipage}{.15\textwidth}
\begin{tikzpicture}[>=stealth',level/.style={sibling distance = 1cm/#1,
  level distance = 0.5cm}] 
\node [bt_n] {}
    child{ node [bt_n] {} 
	child { node [bt_l] {}}
	child { node [bt_l] {}}
	} 
    child{ node [bt_n] {} 
	child { node [bt_l] {}}
	child { node [bt_l] {}}
	}
; 
\end{tikzpicture}
\end{minipage}
\begin{minipage}{.15\textwidth}
\begin{tikzpicture}[>=stealth',level/.style={sibling distance = 1cm/#1,
  level distance = 0.5cm}] 
\node [bt_n] {}
    child{ node [bt_l] {} }
    child{ node [bt_n] {} 
	child{ node [bt_n] {}
		child { node [bt_l] {}}
		child { node [bt_l] {}}
	}
	child{ node [bt_l] {}}
	} 
; 
\end{tikzpicture}
\end{minipage}
\caption{Number of binary trees for $n\in\{0,1,2,3\}$ internal nodes ($n$ goes from $0$ for the top row to $3$ for the bottom row). The black squares correspond to the internal nodes, while the white triangles correspond to the leaves.}\label{fig:catalan}
\end{figure}
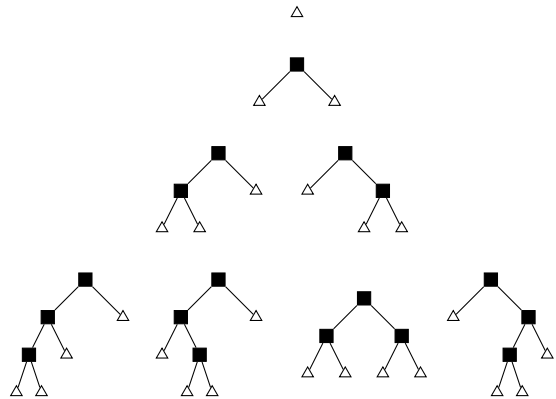

We start by considering the case of binary trees with $n$ internal nodes and $n+1$ leaves. 
The number of such binary trees is $C(n)$, the so-called Catalan number, given by
\begin{equation}
C(n) = \frac{1}{n+1} \binom{2n}{n}. 
\end{equation}
See, for example, \cite{Knuthfasc4}. 
Figure \ref{fig:catalan} shows the $C(0), C(1), C(2)$ and $C(3)$ trees corresponding to $n=0, 1, 2$ and $3$ internal nodes. 

The $n+1$ leaf labels are selected from $P$. 
Each primitive in $P$ can be selected more than once. 
Since there are $n+1$ leaves, there are $|P|^{n+1}$ possible leaf label configurations.\footnote{$|S|$ is the cardinality of the set $S$.} 
\\
The labels for the $n$ inner nodes (operations) are selected from $O$. 
Each operation in $O$ can be selected more than once.
So, there are $|O|^{n}$ possible operation node label configurations.
\\
Thus, in total there are 
\begin{equation}\label{eq:tree_enum}
|P|^{n+1} \cdot |O|^{n} \cdot C(n)
\end{equation}
possible CSG trees with $2n+1$ nodes ($n$ internal nodes and $n+1$ leaf nodes), corresponding to the set of primitives $P$ and the set of Boolean operations $O$.

For a given number of inner nodes $n$, the number of CSG trees is given by (\ref{eq:tree_enum}). 
However, in general, we do not know which value to use for $n$. The only available information is the number of fitted primitives in the set $P$. 
\\
Instead, we use heuristics to find a lower and upper bound for $n$. Let $n_{\min}$ and $n_{\max}$ the minimum and maximum numbers of inner nodes. 
In order to get all possible trees for a given set of primitives $P$, we need to count all possible trees for all possible number of inner nodes between $n_{\min}$ and $n_{\max}$
\begin{equation}\label{eq:sum_tree}
\sum_{i=n_{\min}}^{n_{\max}} |P|^{i+1} \cdot |O|^{i} \cdot C(i). 
\end{equation}

For a given primitive set $P$, we use the following heuristics to estimate $n_{\min}$ and $n_{\max}$. 
We have $n_{\min} = |P|-1$, since the tree should contain all primitives at least once (we do not consider the case where $P$ contains redundant or spurious primitives). Thus there should be at least $|P|$ leaves, resulting in at least $|P|-1$ inner nodes.
\\ 
Strictly speaking, it is not possible to derive a value for $n_{\max}$, since the size of the CSG tree is unbounded. Indeed, it is always possible to add inner nodes and leaves (redundancies) without modifying the geometric set corresponding to the CSG expression (for example by taking the union of the expression with itself).
\\
Instead, we need to look for possible empirical values for $n_{\max}$.
In \cite{friedrich2018}, the estimation for the maximum tree height $h_{\max}\approx \sqrt{\pi/2 \cdot |P|\cdot(|P|-1)}$ is used ($n_{\max}$ can then be estimated from $h_{\max}$). Experiments revealed that it often produces too high values. A tighter choice of $n_{\max}$ depends on the size of the primitive set $P$, the spatial configuration of the primitives as expressed in the intersection graph (see Fig.\,\ref{fig:example1} for an example) and the overall complexity of the model surface that should be represented by the CSG tree. The latter is difficult to quantify in practice.

In the following, we look at possible techniques for reducing the size of the search space and thus simplifying the CSG tree extraction problem. 

\section{Fundamental Products and Disjunctive Normal Form}
\label{sec:dnf}
\subsection{The Full Disjunctive Normal Form}
Similar to the full Disjunctive Normal Form (DNF) for Boolean functions, one can restrict the tree topology to the set of all primitive (or their complement) intersections (the so-called fundamental products \cite{shapiro1991construction}) that are combined via the set union operation
\begin{equation}\label{eq:dnf}
\Phi(P) = \bigcup_{k=1}^{2^{|P|}-1} \epsilon_k \left( g_1 \cap^* g_2 \dots \cap^* g_{|P|} \right), \qquad g_i \in \{p_i, \setminus^{*}p_i\}, 
\end{equation}
where $\epsilon_k$ is equal to one if the corresponding $k-$th fundamental product is included in the CSG expression, and zero otherwise. 
The result is commonly referred to as a two-level CSG representation \cite{shapiro1991construction,Shapiro1991}. 
\\
This formulation reduces the search space complexity to $\mathcal{O}(2^{|P|})$ since there are $2^{|P|}$ fundamental products and we only need to check for each of them if it is inside the target solid $S$. 

A downside of this approach is the excessive size of the resulting CSG expression, since each clause involves the intersection of all the primitives (or their complement). 
\\
When working with Boolean functions, the equivalent of fundamental products that are fully inside the target solid $S$ are called implicants. An implicant that can't be further factored by removing literals is called a prime implicant \cite{Knuthfasc1}. Computing the prime implicants for the CSG expression (\ref{eq:dnf}) can result in a more optimized (more compact) CSG expression. 

\subsection{Non-empty Fundamental Products}
\begin{figure}[!htbp]
	\centering
	\begin{tabular}[c]{cc}
		\begin{subfigure}[c]{0.35\linewidth}
			\includegraphics[width=\textwidth]{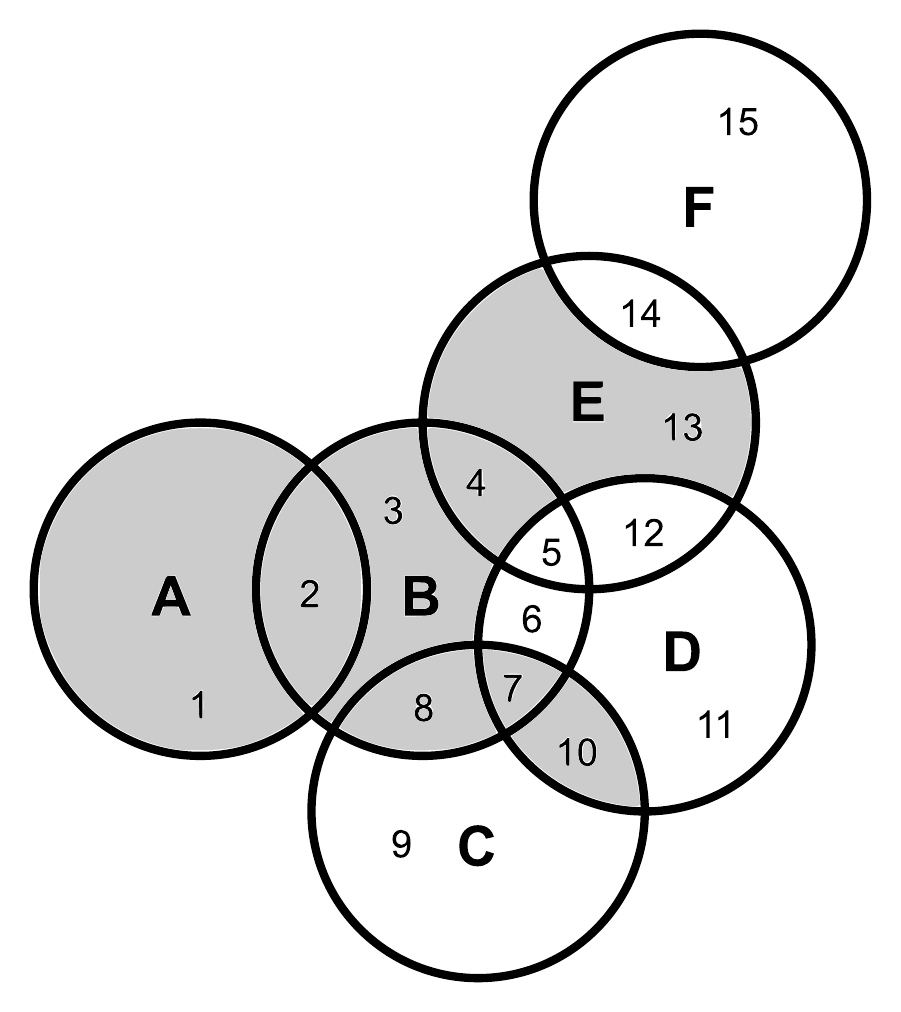}
			\caption{ }
			\label{fig:example0}
		\end{subfigure}&
		\begin{subfigure}[c]{0.5\linewidth}
			\includegraphics[width=\textwidth]{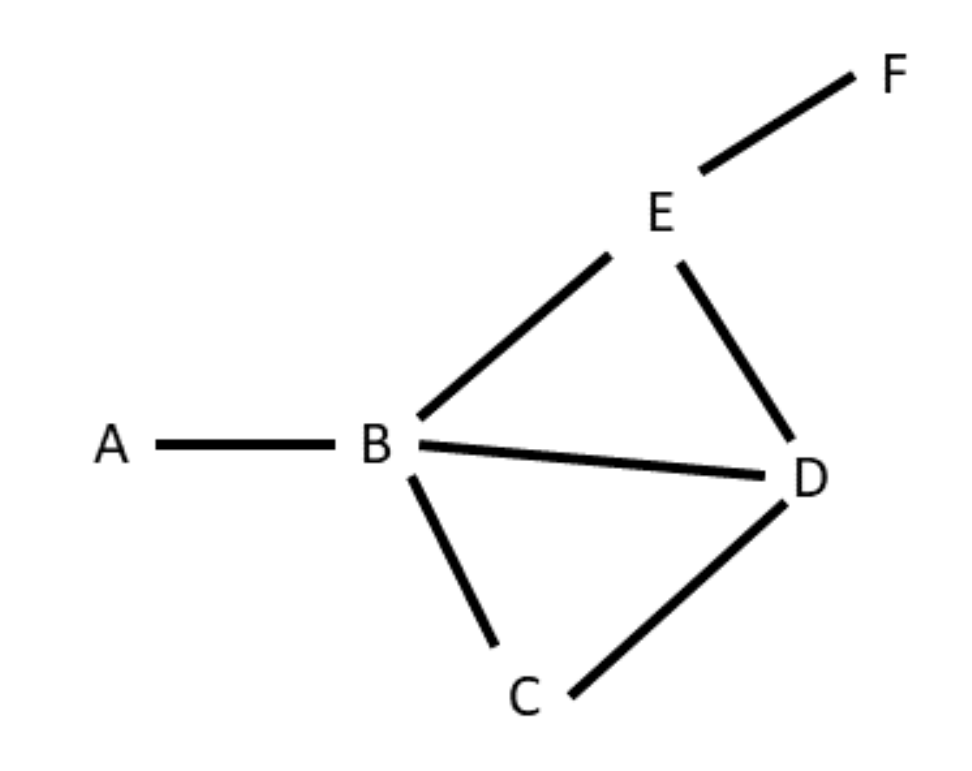}			
			\caption{ }
			\label{fig:example1}
		\end{subfigure}	
	\end{tabular}
	\caption{(a) Example with the primitives $P=\{A,B,C,D,E,F\}$ and $S$, the solid to represent, in grey. The numbers $1$-$15$ identify the non-empty fundamental products. (b) The corresponding intersection graph $G$. The example is adapted from \cite{feld2018optimizing}.}
\label{fig:example}
\end{figure}

The complexity of the search space can be further reduced by noticing that we do not need to consider all the fundamental products, but only those corresponding to non-empty point-sets. See Fig.\,\ref{fig:example0} for an example with a set of primitives and the corresponding non-empty fundamental products. 

The non-empty fundamental products can be determined from the intersection graph $G=(P,E)$ of the primitives in $P$. 
The set of vertices in $G$ corresponds to the set of primitives $P$. There is an edge $(p_i, p_j)$, for $i,j \in \{1, \ldots, |P|\}$, between two vertices $p_i$ and $p_j$ if the corresponding geometric primitives intersect. Figure \ref{fig:example1} shows an example of intersection graph corresponding to the set of primitives shown in Fig.\,\ref{fig:example0}. 
Computing the intersection graph $G$ has a complexity of $\mathcal{O}(|P|^2)$ in the worst case, but can be improved in practice with the use of spatial acceleration structures \cite{zomorodian12fast}. 

When only the non-empty fundamental products are considered, the complexity of the search space becomes proportional to the number of non-empty fundamental products $n_{f}$. 
If the geometric primitives in $P$ are all spatially disjoints, then $E$ is empty and $n_f$ reaches its minimum value, $n_f=|P|$. 
If $G$ is fully connected, then $n_{f}$ reaches its maximum value, $n_f=2^{|P|}-1$. 
Please note, that in general it is not possible to decide whether a fundamental product is empty or not by just considering the intersection graph since it depends on the particular shape of the primitives involved.

This approach still results in possibly large CSG expressions. 
A better method to further reduce the search space and keeping the tree size limited is described in the next section.

\section{Dominant Halfspaces and Solid Decomposition}
\label{sec:decomp}
Dominant halfspaces $\{d_1,...,d_n\} \subseteq P$ are primitives that are located either fully inside or fully outside of the target solid $S$. 
For example, primitives $A$ and $F$ in Fig.\,\ref{fig:example0} are dominant primitives of the solid in grey. 

A solid can be decomposed using dominant halfspaces as \cite{shapiro1991construction} 
\begin{equation}
\label{eq:dec}
S = ((...( S_{rem} \circ d_1 ) \circ ...) \circ d_2) \circ d_n,
\end{equation}
where $S_{rem}$ is the remaining solid after decomposition and $\circ$ is either the difference operator if the following primitive in the expression dominates $\setminus^{*}S$ or the union operator if it dominates $S$. The remaining solid $S_{rem}$ can be described as an expression containing all the remaining non-dominant primitives. 

The time complexity of the decomposition algorithm is $\mathcal{O}(|P|^2)$.
In the worst case, each iteration of the decomposition results in a single primitive being removed from $S$. Thus, the first iteration visits each primitive once ($|P|$ visits) to check if it is dominant.
After removing one single dominant primitive, the second iteration needs $|P|-1$ visits, and so on, resulting in $\sum_{k=|P|}^{1}k = \frac{|P|^2 + |P|}{2}$ necessary visits in total. 

The decomposition can be applied recursively, making it a powerful tool for search space reduction. 
Furthermore, the expression is optimal, since each dominant halfspace is used exactly once in the output expression \cite{shapiro1991construction}. 
Early factoring of dominant halfspaces is used in the following approaches for BRep to CSG conversion \cite{Buchele1999,Buchele2003,buchele2004three}. 

If $S_{rem}$ is not empty after the decomposition, one needs to compute a CSG expression for the remaining solid from the remaining non-dominant primitives. 
For example, one can use the approach described in Section~\ref{sec:dnf}, and build the intersection graph of the non-dominant primitives. 
For sufficiently large models, this graph is not connected and a connected component analysis results in a set of sub-intersection graphs. 
The corresponding expressions for each sub-graph can be extracted independently and the result is then merged. 
This can be used to further reduce the search space  (see, for example, \cite{friedrich_gecco2019,friedrich2020csg-optim}).

\bibliographystyle{plain}
\bibliography{references}

\end{document}